\documentclass[manuscript=article,layout=twocolumn]{achemso}
\usepackage[version=3]{mhchem} 
\usepackage{amsmath,amssymb}
\usepackage{graphicx}
\usepackage[utf8]{inputenc}
\usepackage{pdflscape}
\usepackage{epstopdf}  
\usepackage{dcolumn}   
\usepackage{bm}        
\usepackage{times}
\usepackage{empheq}
\usepackage{tabularx}
\newcolumntype{V}{>{\hsize=1.2\hsize\centering  \arraybackslash}X}
\newcolumntype{b}{>{                \raggedleft \arraybackslash}X}
\newcolumntype{B}{>{                \centering  \arraybackslash}X}
\newcolumntype{m}{>{\hsize=0.7\hsize\raggedleft \arraybackslash}X}
\newcolumntype{M}{>{\hsize=0.7\hsize\centering  \arraybackslash}X}
\newcolumntype{s}{>{\hsize=0.5\hsize\raggedleft \arraybackslash}X}
\newcolumntype{S}{>{\hsize=0.5\hsize\centering  \arraybackslash}X}

\newcommand{\lr}       {\ensuremath{^\text{lr}}}
\newcommand{\sr}       {\ensuremath{^\text{sr}}}
\newcommand{\lrx}      {\ensuremath{^{\text{lr}(x)}}}
\newcommand{\srx}      {\ensuremath{^{\text{sr}(x)}}}
\newcommand{\RSH}      {\ensuremath{_\text{RSH}}}
\newcommand{\RPA}      {\ensuremath{_\text{RPA}}}
\newcommand{\Hxc}      {\ensuremath{_\text{Hxc}}}
\newcommand{\T}[1]     {\ensuremath{\text{#1}}}
\newcommand{\DC}       {\ensuremath{\ensuremath{\mathbf{\Delta}}_\text{DC}}}

\newcommand{\trace}[1] {\ensuremath\text{tr}\left\{ #1 \right\}}
\newcommand{\bra}[1]   {\ensuremath{\langle #1 \vert}}
\newcommand{\ket}[1]   {\ensuremath{\vert #1  \rangle}}

\newcommand{\trnp}     {\ensuremath\T{T}}

\newcommand{\zero}     {^{(0)}}
\newcommand{\deux}     {^{(2)}}
\newcommand{\x}        {^{(x)}}
\newcommand\FdTr[2]    {\left\{#1,#2\right\}}
\renewcommand{\ll}      {\ensuremath{\lambda}}
\renewcommand{\b}[1]   {\ensuremath{\mathbf{#1}}}

\newcommand\B[1]       {{\bm #1}}
\newcommand\derivR[2]    {\frac{\partial #1}{\partial #2}}
\newcommand\half         {\frac{1}{2}}
\newcommand\Shalf        {{\textstyle\frac{1}{2}}}
\newcommand\fct[3]{\mathcal{X}_#1^{#2}\!\left(#3\right)}

\author{Bastien Mussard}
\affiliation[Université de Lorraine]
{CRM2, Institut Jean Barriol, Universit\'e de Lorraine, F-54506 Vand{\oe}uvre-l\`{e}s-Nancy, France}
\author{P\'eter G.\ Szalay}
\affiliation[Eötvös University]{Institute of Chemistry, E\"otv\"os Loránd University, H-1518 Budapest, P.O.Box. 32, Hungary}
\author{J\'anos G. \'Angy\'an}
\email{janos.angyan@univ-lorraine.fr}
\phone{+33-(0)3 83 68 48 74}
\fax{+33-(0)3 83 68 43 00}
\affiliation[Université de Lorraine]
{CRM2, Institut Jean Barriol, Universit\'e de Lorraine, F-54506 Vand{\oe}uvre-l\`{e}s-Nancy, France}
\alsoaffiliation[CNRS]
{CRM2, Institut Jean Barriol, CNRS, F-54506 Vand{\oe}uvre-l\`{e}s-Nancy, France}

\title[Analytical RSH+RPA gradients]
{Analytical Energy Gradients in Range-Separated Hybrid Density Functional Theory with Random Phase Approximation}
\begin{document}
\begin{abstract} 
Analytical forces have been derived in the Lagrangian framework for several random phase approximation (RPA) correlated total energy methods based on the range separated hybrid (RSH) approach, which combines a short-range density functional approximation for the short-range exchange-correlation energy with a Hartree-Fock-type long-range exchange and RPA long-range correlation. The RPA correlation energy has been expressed as a ring coupled cluster doubles (rCCD) theory. The resulting analytical gradients have been implemented and tested for geometry optimization of simple molecules and intermolecular charge transfer complexes, where intermolecular interactions are expected to have a non-negligible effect even on geometrical parameters of the monomers. 
\end{abstract}
\section{Introduction}

After having achieved the limit of chemical accuracy with the hybrid functionals for a vast majority of applications, density functional theory (DFT) in its Kohn-Sham~\cite{Kohn:65}, or more precisely generalized Kohn-Sham~\cite{Kuemmel:07} formulation became the most popular electronic structure method of computational chemistry~\cite{Burke:12} and computational material sciences~\cite{Hafner:08}. In the mean time, new applications revealed more and more examples where routine DFT methods fail  and pointed out some inherent weaknesses of the usual local and semi-local approximation of the exchange-correlation functionals~\cite{EngelDreizler:book}. During the last years, considerable progress has been achieved in a better understanding of the fundamental reasons of these failures~\cite{Perdew:09}, e.g.\ the lack of a proper description of London dispersion forces or the systematic errors due to the erroneous treatment of charge and spin localization/delocalization~\cite{Cohen:12a}. Various methods have been proposed to remedy functionals for one or more of the above-mentioned problems. Among these approaches, methods that use both occupied and virtual orbitals, sometimes called methods belonging to the 5th rung of Jacob's ladder~\cite{Perdew:01f}, play a privileged role due their extraordinary flexibility, limited only by the associated computational costs. Simple second order Rayleigh-Schr\"odinger perturbation theory  can be an acceptable solution for many cases, but one runs into serious difficulties in small-gap systems. The next level of approximation, which has a long-standing tradition in DFT~\cite{Langreth:75}, is the random phase approximation (RPA). The RPA can be derived either in an adiabatic-connection/fluctuation-dissipation-theorem framework~\cite{Hesselmann:11x,Eshuis:12b} or as a particular approximation to the coupled cluster doubles (CCD) expression of the correlation energy in diagrammatic perturbation theory~\cite{Scuseria:08}. The two types of formulations are intimately related and in some special cases they lead to correlation energies that are strictly equivalent~\cite{Jansen:10}. The great advantage of RPA is that it provides a correlation energy functional, which is compatible with Hartree-Fock exchange.

Random phase approximation (RPA) methods, belonging to the 5th rung of Jacob's ladder~\cite{Perdew:01f} of DFT approaches, are becoming a practical tool to construct correlation energy functionals~\cite{Furche:01b,Fuchs:02,Miyake:02,Furche:05,Fuchs:05,Furche:08,Harl:08,Lu:09,LiYan:10,Gruneis:09,Nguyen:09,Nguyen:10,Paier:10,Hesselmann:10,Hesselmann:11x,Eshuis:12b,Hesselmann:12,Ren:12,Toulouse:09,Janesko:09,Janesko:09a} not only in computational material sciences but also in the quantum chemistry applications. Recent works by Eshuis and Furche~\cite{Eshuis:10,Eshuis:12b}, Hesselmann~\cite{Hesselmann:12}, Ren \textit{et al.}~\cite{Ren:12} and others~\cite{Lu:09,Harl:08} have demonstrated that highly efficient RPA implementations with favorable scaling properties with the system size are conceivable. Thus, it is now becoming possible to study extended systems~\cite{Eshuis:10,Hesselmann:12} with similar computational resources but with a considerably better accuracy than the MP2 method. Quite good results could be obtained for reaction energies and barriers~\cite{Paier:12} and especially good performance is expected for systems where van der Waals interactions~\cite{Bjorkman:12}, and in particular London dispersion forces~\cite{Eshuis:11}, play a crucial role.  

A few shortcomings of the RPA method have also been identified since the first numerical studies on molecular systems~\cite{Furche:01b,Fuchs:02,Miyake:02} and solids~\cite{Marini:06,Harl:08,Nguyen:09,Lu:09}. As it has been expected from earlier work of Perdew and his colleagues~\cite{Yan:00}, the RPA describes rather poorly the short-range correlation: for this reason, e.g.\ its performance for atomization energies is inferior to that of a good gradient corrected functional~\cite{Furche:01b}. Another problem is due to the slow basis set convergence of the RPA correlation energy, which makes the determination of CBS (complete basis set) limit correlation energies (and energy differences) for larger systems prohibitively expensive~\cite{Furche:05,Eshuis:12a}. These problems can be fixed to a considerable extent by applying instead of the previously mentioned "full-range" RPA techniques, which use Kohn-Sham orbitals obtained by standard (usually PBE) exchange-correlation functionals, followed by the frozen-orbital evaluation of the Hartree-Fock (HF) exchange and the RPA correlation, the combination of a short-range DFA (density functional approximation) and a long-range RPA method within the range-separated hybrid (RSH) framework~\cite{Toulouse:09,Toulouse:10a,Janesko:09,Janesko:09a,Janesko:09b,Janesko:09c}.

This family of methodologies consists in separating the long- and short-range electron-electron interactions in the Hamiltonian. The short range interactions are described within a density functional approximation (DFA), using appropriately designed short-range LDA or PBE exchange-correlation functionals. The long-range exchange is taken into account by a nonlocal Hartree-Fock operator, while the long-range correlation is described in the random phase approximation. 
Range-hybrid RPA methods have shown quite good performance for intermolecular complexes, especially when certain RPA variants are applied for the long-range correlation~\cite{Toulouse:11}. In particular, the stacking interactions, which are due to long-range dynamical correlation effects, are well described. Considering the nature of the RPA, it is expected that its best theoretical performance is deployed for long-range dynamical correlations, which are physically responsible for London dispersion forces. Since the reference determinant for the long-range RPA calculations is constructed from RSH orbitals, which are optimized using a long-range HF exchange potential, spurious delocalization effects,  common for conventional local and semi-local functionals can be avoided from the outset. In fact, the presence of the long-range HF-exchange reduces considerably the delocalization error, which usually deteriorates the description of charge transfer (donor-acceptor) complexes. 
 Furthermore, since RSH orbitals fulfill a long-range Brillouin-theorem~\cite{Angyan:05}, there is no need for single-excitation corrections, as proposed by Ren and his coworkers~\cite{Ren:11}. It has been shown that the basis set convergence properties of the range-hybrid RPA method is much faster than that of the "full-range" correlation methods and the basis set superposition error has an almost negligible impact on the results even at relatively small basis sets~\cite{Zhu:10}.  Although actual computational implementations of RSH+RPA approaches are far from being optimal, this class of methods remains one of the most promising ways to correct a number of notorious shortcomings of conventional DFAs.

We should be aware of the fact that the efficient and accurate calculation of total energies is not enough to perform realistic modeling work: one should definitely go beyond  total energies and in order to relax  atomic positions (geometry optimization) or  follow the evolution of the system by solving the equations of motion for the nuclei (Born-Oppenheimer molecular dynamics) one needs the corresponding analytical total energy derivatives (gradients) in each point of the potential energy surface. 
Since the pioneering work of Pulay on analytical Hartree-Fock derivatives~\cite{Pulay:69}, which opened the way to a routine analytical calculation of forces,  numerous theoretical and computational developments have been proposed, making analytical gradients accessible for a vast majority of mainstream electronic structure methods. The formal difficulties which appeared at the beginning for non-variational energy expressions have been overcome by the Lagrangian formulation of derivative properties~\cite{Helgaker:89,Helgaker:89a}, permitting to have analytical forces in  ground and excited states~\cite{Szalay:95}. Two very recent publications on RPA analytical gradients, which appeared after  completing the present work,  have been based also on the Lagrangian method.  Analytical gradients of post-Hartree-Fock RPA correlation energies have been published by Helgaker's group~\cite{Rekkedal:13a} in a ring coupled cluster type formulation of RPA~\cite{Scuseria:08}, while  Furche and his coworkers~\cite{Burow:13} used a direct RPA correlation energy expression, formulated via the frequency dependent dielectric matrix in the resolution of identity approximation, using Kohn-Sham orbitals.  Our approach, developed independently from theirs, follows a Lagrangian strategy too. 

Our main objective was to derive RSH+RPA gradients, i.e.\ the analytical first derivative of the combination of a short-range DFA  and a long-range RPA method within the range-separated hybrid (RSH) framework. We propose gradient expressions for different variants of the long-range RPA correlation energy~\cite{Angyan:11,Toulouse:11}, without and with exchange. Our expressions, at the extreme limits (zero and infinity) of the range-separation parameter, become identical either with post-Hartree-Fock RPA, or with pure DFA analytical gradients. The second-order limit of the RSH+RPA correlation energy with exchange is the RSH+MP2~\cite{Angyan:05}. For this latter case, more precisely for the long-range local MP2 approach~\cite{Goll:08,Goll:08a} combined with short-range DFT, analytical gradients are already available~\cite{Chabbal:10}, and implemented in the MOLPRO program suite~\cite{MOLPRO:12}. 

In the first subsection of Section~II, we provide a quick overview of the range separated hybrid + random phase approximation (RSH+RPA) method. Our derivation of the gradient of the RSH+RPA total energy is applicable for numerous variants of the long-range RPA correlation energy: each of these variants will be shortly discussed. The second subsection explains the construction of the Lagrangian for a generic RSH+RPA method and, in the third subsection, we present explicit working expressions for the gradients. Section~III provides some practical details about our implementation, followed by illustrative applications on geometry optimizations of small molecules and of charge transfer complexes. The paper is closed by  conclusions and a short outlook of future developments. We use atomic units throughout the whole paper. 


\section{Theory}

\subsection{RSH energy}
\label{sec:rsh}

In the range-separated hybrid DFT framework one starts with a self-consistent independent-particle calculation by minimizing the energy
\begin{equation}
\label{eq:RSHenergy}
E\RSH      = \min_{\Phi} \left\{ \bra{ \Phi} \hat{T} + \hat{V}_\T{ne} + \hat{V}\lr_{ee} \ket{\Phi} + E\sr\Hxc[n_{\Phi}] \right\}, 
\end{equation}
with respect to the molecular orbitals $\phi_i(\B{r})$ of the single determinant $\Phi$. 
Here 
$n_{\Phi}$ is the density associated with the single-determinant wave function, 
$\hat{T}$ is the kinetic energy operator, 
$\hat{V}_\T{ne}$ is the nuclear attraction operator, 
$\hat{V}_{ee}\lr$ is a long-range electron-electron repulsion operator constructed with the error function $v\lr_{ee}(r) = \T{erf}(\mu r)/r$, 
and 
$E\Hxc\sr[n_\Phi]$ is the associated short-range Hartree-exchange-correlation 
density functional, written  as:
\begin{equation}
E\Hxc\sr[n_\Phi]=\int d\B{r}\; F\bigl(\B{\xi}(\B{r})\bigr),
\label{eq:Hxc}
\end{equation}
where $\B{\xi}(\B{r})$ is an array of quantities such as the density 
\begin{align}
n(\B{r}) &= \sum_i\,      d_{ii}    \zero \,\phi_i^\ast\!(\B{r})\,\phi_i(\B{r})
\nonumber\\
         &= \sum_{\mu\nu} D_{\mu\nu}\zero \,\chi_\mu^\ast(\B{r})\,\chi_\nu(\B{r}),
\end{align}
and other density-related parameters (spin density, reduced density gradients, 
\textit{etc}\dots) entering in the definition of the functional. 
$\xi_A$ will designate a component of $\B{\xi}$,  
$\b{d}\zero$ (\textit{resp.} $\b{D}\zero$) is the RSH density matrix in the molecular orbital basis (\textit{resp.} in the atomic orbital basis). 
Throughout the paper, the indices $p$, $q$, $r$, $s$ are used for general molecular orbitals, $i$, $j$, $k$ for occupied molecular orbitals and $a$, $b$, $c$ for virtual molecular orbitals; the indices $\mu$, $\nu$, $\rho$, $\sigma$, are used for atomic orbitals.

It will be convenient to express the RSH energy with a fockian $\b{f}=\b{h} + \b{g}\sr + \b{g}\lr$, where the short- and long-range two-electron contributions are:

\begin{subequations}
\begin{empheq}{align}
g\sr_{pq}&=\sum_A \int d\B{r}\; \derivR{F}{\xi_A(\B{r})}\derivR{\xi_A(\B{r})}{d\zero_{pq}}\\
g\lr_{pq}&\doteq   g\lr\left[\b{d}\zero\right]_{pq}
\nonumber\\&=\sum_{rs}d\zero_{rs}\left( ( pq| rs )\lr -\Shalf ( pr| qs )\lr \right),
\end{empheq}
\end{subequations}
and the two-electron integrals follow the chemist's notation. 

Using the above-defined quantities, the RSH energy of Eq.~(\ref{eq:RSHenergy}) reads as:
\begin{align}
E\RSH      & = \trace{\b{d}\zero\b{f}} - 
\half\trace{\b{d}\zero\b{g}\lr}
\nonumber \\ &\quad + 
\biggl(E\Hxc\sr[n_0]-\trace{\b{d}\zero\b{g}\sr}\biggr),
\end{align}
where the last terms are the double-count correction of the long-range Hartree-Fock energy 
($\DC\lr \!\!=\!\!-\half\trace{\b{d}\zero\b{g}\lr}$) on the one hand,
and the sum of the short-range Hartree 
and exchange-correlation energies ($\DC\sr = E\Hxc\sr[n_0] - \trace{\b{d}\zero\b{g}\sr}$), on the other hand. 

\subsection{RPA energy}
\label{sec:rpa}

Since the minimizing RSH wave function
is a single-determinant approximation to the exact wave function, the long-range part of the RSH energy contains only Hartree and exchange terms. 
In principle, the exact ground-state energy can be obtained from the RSH energy by adding the long-range correlation energy $E_c\lr$:
\begin{equation}
E = E\RSH      + E_c\lr,
\label{eq:Etot}
\end{equation}
provided the exact short-range exchange-correlation functional were known and we could solve exactly the long-range correlation problem. 
In this work we are interested in the random phase approximation (RPA) to the long-range correlation energy $E_c\lr$. As it has been discussed in previous works~\cite{Angyan:11,Toulouse:11}, several alternative RPA variants exist, and most of them can be expressed either in an adiabatic-connection formalism~\cite{Angyan:11}, or can be reformulated as ring approximations in the coupled cluster doubles (CCD) framework~\cite{Scuseria:08}. Among the numerous variants proposed in the CCD framework~\cite{Toulouse:11}, we focus our attention to the following ones:
\begin{enumerate}
\item Direct RPA (dRPA-I)
\begin{equation}
E_{c,\T{dRPA-I}}\lr = \half\trace{ {}^1\b{K}\lr \; {}^1\b{T}_\T{dRPA}\lr },
\end{equation}
with the dRPA amplitudes $^1\b{T}_\T{dRPA}\lr$ satisfying the dRPA Riccati equations:
\begin{multline}
\label{eq:dRPARicatti}
\b{R}_\T{dRPA}=  (\b{I}+ {}^1\b{T}_\T{dRPA}\lr) \, {}^1\b{K}\lr \, (\b{I} + {}^1\b{T}_\T{dRPA}\lr) \\
+ {}^1\b{T}_\T{dRPA}\lr \, \B{\epsilon} 
+ \B{\epsilon} \, {}^1\b{T}_\T{dRPA}\lr
=\b{0}
\end{multline}

\item Direct RPA with SOSEX (SOSEX)
\begin{equation}
E_{c,\T{SOSEX}}\lr = \half\trace{ {}^1\b{B}\lr \; {}^1\b{T}_\T{dRPA}\lr },
\end{equation}
where the amplitudes satisfy the previous dRPA Riccati equations, Eq.~(\ref{eq:dRPARicatti}). 
\item Approximate exchange RPA (RPAX2)
\begin{equation}
E_{c,\T{RPAX2}}\lr = \half\trace{ {}^1\b{K}\lr \; {}^1\b{T}_\T{RPAX}\lr },
\end{equation}
with the RPAX ${}^1\b{T}_\T{RPAX}\lr$ amplitudes satisfying the following Riccati-like equation:
\begin{multline}
\b{R}_\T{RPAX}=   (\b{I}+ {}^1\b{T}_\T{RPAX}\lr) \, {}^1\b{B}\lr \, (\b{I} + {}^1\b{T}_\T{RPAX}\lr) \\
+  {}^1\b{T}_\T{RPAX}\lr \, \B{\epsilon}
+  \B{\epsilon} \, {}^1\b{T}_\T{RPAX}\lr
=\b{0}.
\end{multline}
This last variant has been suggested by Hesselmann in a rCCD formalism~\cite{Hesselmann:12}. An analogous approach can be derived in an adiabatic connection framework as well~\cite{Jansen-ms:13}. 

\end{enumerate}   

The above equations are expressed in terms of symmetry adapted super-matrices for a closed shell system. The matrix elements of $\B{\epsilon}$ are $\epsilon_{ia,jb}=f_{ab}\delta_{ij}-f_{ij}\delta_{ab}$, ${^1\b{K}}\lr$ and ${^1\b{B}}\lr$ are combinations of two-electron integrals over spatial molecular orbitals:
\begin{align}
^1{K}_{ia,jb}\lr &= 2 (ia | jb )\lr = 2{K}_{ia,jb}\lr
\\
^1{B}_{ia,jb}\lr &= 2 (ia | jb )\lr - (ib | ja )\lr = 2{K}_{ia,jb}\lr - {K}^{\prime\text{lr}}_{ia,jb}. 
\end{align}

\subsection{RSH+RPA Lagrangian}
\label{sec:lagrangian}

Since the RPA correlation energy is non-variational, we use the Lagrangian formalism to evaluate the RSH+RPA gradients. The Lagrange functional is constructed from the total RSH+RPA energy expression of Eq.~(\ref{eq:Etot}) along with a set of undetermined Lagrange multipliers associated with the constraints that the parameters entering the energy expression must fulfill. In our case we have to consider three constraints which ensure that
    (1) the orbitals are solutions of the RSH equations,
    (2) they remain orthogonal
and (3) the amplitudes defining the RPA correlation energy are always solutions of the Riccati-like equations. The Lagrangian then reads:
\begin{multline}
\mathcal{L}(\b{C},\b{T},\b{z},\b{x},\B{\ll})= E\RSH     (\b{C})+E\RPA\lr(\b{C},\b{T})
\\               + \trace{\b{z}\b{f}}
 + \trace{\b{x}\left(\b{C}^\trnp\b{S}\b{C}-\b{1}\right)} \\
  + \trace{\B{\ll}\b{R}(\b{C},\b{T})}, 
\label{eq:lag}
\end{multline} 
where $\b{C}$ is the matrix of the MO coefficients obtained from the self-consistent RSH equations in the LCAO framework, $\b{T}$ is the super-matrix of the RPA amplitudes in MOs, $\b{z}$ is the set of multipliers associated with the Brillouin conditions, $\b{x}$ is the multiplier matrix for the orthogonality conditions and $\B{\ll}$ is the super-matrix related to the Riccati conditions $\b{R}(\b{C},\b{T})$. For the sake of notational simplicity, the the following the superscript "lr" will be omitted for quantities, where there is no risk of confusion (like RPA amplitude matrices). Explicit "lr/sr" labels will be kept only for cases where their use seemed indispensable for understanding. 

The Lagrange multipliers $\B{\ll}$ are obtained from the stationary condition of the Lagrangian with respect to the amplitudes $\b{T}$, which can be written for all the versions of RPA introduced in Section \ref{sec:rpa} as:
\newcommand\mat{\b{Q}(\b{T})}
\begin{align}
\label{eq:lambda}
\mat \B{\lambda}+\B{\lambda} \mat^\trnp=-\b{P},
\end{align}
where the expression of the super-matrix $\mat$ depends on the Riccati equation considered and the right-hand side $\b{P}$ is a combination of two-electron integrals, having a form that depends on the correlation energy expression used for $E\RPA\lr$.  

After some lengthy algebra outlined in the Appendix \ref{appsec:statioC}, one can show that the stationary conditions of the Lagrangian with respect to the orbital coefficients $\b{C}$ (more precisely: with respect to the first-order variation of these coefficients, $\b{V}$) boil down to the following two equations:

\begin{subequations}
\begin{empheq}[left={}\empheqlbrace]{align}
\label{eq:z}
&\Big(
\B{\Theta}-\B{\Theta}^\trnp+\b{f}\b{z}-\b{z}\b{f}
\nonumber\\&
\quad\qquad+4\b{g}\lr\left[\b{z}\right]+4\b{w}\sr[\b{z}]\Big)_{ai}=0\\
&\B{\Theta}+\B{\Theta}^\trnp+\tilde{\B{\Theta}}(\b{z})+\tilde{\B{\Theta}}(\b{z})^\trnp=-4\b{x}
\label{eq:x}
\end{empheq}
\end{subequations}

The first equation is a coupled perturbed RPA equation, which should be solved for $\b{z}$, the difference density matrix in the virtual-occupied block; the second equation gives $\b{x}$, the energy-weighted density matrix, once $\b{z}$ is known. 
The terms that depend neither on $\b{z}$ nor on $\b{x}$ are collected in the matrix $\B{\Theta}$, while $\tilde{\B{\Theta}}(\b{z})$ regroups terms that depend on $\b{z}$.

As it is shown in the Appendix, there is a close correspondence between the procedures used to derive the long- and  short-range contributions to Eq.~(\ref{eq:z}) and Eq.~(\ref{eq:x}) due to the appearance of two analogous terms, denoted by $\b{g}\lr\left[\b{z}\right]$ and $\b{w}\sr[\b{z}]$ respectively. Furthermore, in the expression of the Lagrange multiplier, 
$\b{x}$, defined by Eq.~(\ref{eq:x}) one can identify contributions from the derivatives of both the RPA long-range correlation and the RSH reference energies, \textit{i.e.} one can write for the occupied-occupied block:

\begin{align}
\left(\b{x}\right)_{ij}=\left(\b{x}\RPA\lr\right)_{ij}-2\left(\b{f}     \right)_{ij},
\end{align}
where the second term is the usual expression of the $\b{x}$ multiplier coming from the gradient of the reference energy.

\subsection{Analytical gradients }

Once the equations (\ref{eq:lambda}), (\ref{eq:z}) and (\ref{eq:x}) have been solved for the multipliers $\B{\ll}$, $\b{z}$ and $\b{x}$, the Lagrangian is fully known. By the virtue of its variational property with respect to all of its parameters and since at its minimum it is equal to the RSH+RPA energy, the gradient is given by:

\begin{align}
E                         \x=\mathcal{L}\x&=
\trace{\b{D}^1 \b{H}\x}
+ \trace{\b{X}\b{S}\x} 
+ E_\text{DFT}\srx
\nonumber\\&\quad 
+ \sum_{\mu\nu\rho\sigma}\left(\b{D}^2+\B{\Gamma}^2\right)_{\mu\nu,\rho\sigma}(\mu\nu|\rho\sigma)\lrx,
\label{eq:grad}
\end{align} 
with $\left(\b{X}\right)_{\mu\nu}=\sum_{pq}C_{\mu p}\,(\b{x})_{pq}\,C^\trnp_{q \nu}$, 
and: 

\begin{align}
\left(\b{D}^1\right)_{\mu\nu}
&=\sum_{pq} C_{\mu p}\left(\b{d}\zero+\b{d}\deux+\b{z}\right)_{pq}C^\trnp_{q \nu}
\nonumber\\
&=\left(\b{D}\zero+\b{D}\deux+\b{Z}\right)_{\mu\nu}.
\end{align}

As mentioned previously, the matrix of the Lagrange multipliers, $\b{x}$, can be identified as the energy-weighted one-particle density matrix, while  $\b{z}$ is the difference density matrix. The matrix $\b{d}\deux$ is defined in the Appendix \ref{appsec:statioC}. 
For the sake of clarity, in the four-index relaxed two-particle density
appearing in equation (\ref{eq:grad})  we have separated in  the contributions that appear usually in various types of gradient expressions,
\begin{align}
\left(\b{D}^2\right)_{\mu\nu,\sigma\rho}&=\left(\Shalf\b{D}\zero+\b{D}\deux+\b{Z}\right)_{\mu\nu} D\zero_{\rho\sigma}
\nonumber\\&\qquad
   -\half\left(\Shalf\b{D}\zero+\b{D}\deux+\b{Z}\right)_{\mu\rho}D\zero_{\nu\sigma},
\end{align}
and those contributions, which are specific to the RSH+RPA gradient expression:

\begin{align}
\left(\B{\Gamma}^2\right)_{\mu\nu,\sigma\rho}&=\sum_{ia,jb}C_{\mu j}C_{\nu i}C^\trnp_{a\sigma}C^\trnp_{b\rho}\left(\b{M}\right)_{ia,jb}
\nonumber\\&\qquad
      +\sum_{ia,jb}C_{\mu j}C_{\nu i}C^\trnp_{b\sigma}C^\trnp_{a\rho}\left(\b{N}\right)_{ia,jb}.
\end{align}

The expressions of the super-matrices $\b{M}$ and $\b{N}$ depend on the Riccati equation and long-range correlation energy formula, specific to a given RPA variant.  
These quantities are coming from the factorization of the Lagrangian with respect to terms that depend on the orbital coefficients. 
The gradient expressions of the various RPA versions considered in this paper differ only in the details of the $\b{M}$ and $\b{N}$ super-matrices. 
For example, in the case of dRPA-I their expressions are:

\begin{align}
\b{M} &=\half {}^1\b{T}_\T{dRPA} 
      + \B{\lambda}
      + \B{\lambda} \; {}^1\b{T}_\T{dRPA}
      + {}^1\b{T}_\T{dRPA} \; \B{\lambda}
\nonumber\\&\quad
      + {}^1\b{T}_\T{dRPA} \; \B{\lambda} \; {}^1\b{T}_\T{dRPA}
\\
\b{N} &=\b{0}.
\end{align}

The derivative of the density functional part with the corresponding double-counting term reads, on a real-space grid of points $\{\lambda\}$ which have the integration weights $\omega_\lambda$:

\begin{align}
\label{eq:gradsrdft}
E_\text{DFT}\srx& =
  \sum_A\left\{\sum_\lambda
  \omega_\lambda  \left(F(\xi_{A,\lambda}) + \derivR{F}{\xi_A}\left(\xi_{A,\lambda}^{\b{d}\deux   }+\xi_{A,\lambda}^ \b{z}    \right)\right)
       \right\}\x
\nonumber\\&
= \sum_\lambda \sum_A 
  \omega_\lambda\x\left(F(\xi_{A,\lambda}) + \derivR{F}{\xi_A}\left(\xi_{A,\lambda}^{\b{d}\deux   }+\xi_{A,\lambda}^ \b{z}    \right)\right)
\nonumber\\&\quad
+ \sum_\lambda \sum_A 
  \omega_\lambda        \derivR{F   }{\xi_A             }\left(\xi_{A,\lambda} ^{\b{d}\zero(x)}+\xi_{A,\lambda}^{\b{d}\deux(x)}+\xi_{A,\lambda}^{\b{z}(x)}\right)
\nonumber\\&\quad
+ \sum_\lambda \sum_{AB}
  \omega_\lambda        \derivR{^2 F}{\xi_B\partial\xi_A}\left(                                 \xi_{A,\lambda}^{\b{d}\deux   }+\xi_{B,\lambda}^ \b{z}    \right)\xi_{B,\lambda}\x.
\end{align}

Further details about the short-range DFT gradient contributions can be found in Appendix \ref{appsec:derivSR}. Note that in the limiting case of vanishing range-separation parameter, $\mu$, one obtains the full-range DFA gradients, for
$\mu \rightarrow \infty$, one gets the Hartree-Fock + RPA gradients, as given by Rekkedal \textit{et al.}~\cite{Rekkedal:13a}, and by omitting completely the long-range RPA at any finite value of $\mu$, one gets simply the analytical gradient expression of the RSH energy.

\section{Tests and applications}

The above equations have been implemented in the development version of MOLPRO~\cite{MOLPRO:12}. Taking advantage of some similarities in the structure of the RSH+RPA and RSH+MP2 gradients, our implementation closely follows the flowchart of the MOLPRO MP2 gradient program~\cite{Schutz:04} previously adapted for the range-hybrid methods~\cite{Chabbal:10}. 
Computational times are as expected, i.e.\ at the sr-LDA+lr-RPA level there is the same percentage difference between the timings of the energy and gradient calculations as at the sr-LDA+lr-MP2 level. 
                    There is also the same percentage additional cost when 
                    comparing an sr-LDA+lr-MP2 calculation to an sr-LDA+lr-RPA calculation of either the energy or the gradients. 
The scaling of the gradient calculation follows that of an energy calculation (in the present implementation $N^6$). However the algorithm
can take easily advantage of future density-fitting implementations, which can be  as low as $N^4$ for the dRPA-I method and $N^5$ for other RPA variants.
 
For all the data presented here, we calculate the mean absolute error (MAE), defined as $\tfrac{1}{N}\sum_i |a_i-a^\T{ref}_i|$, and the mean percentage absolute error defined as $\tfrac{1}{N}\sum_i |\tfrac{a_i-a^\T{ref}_i}{a^\T{ref}_i}|$.

\begin{figure}[!htbp]
\begin{center}
\includegraphics[width=\linewidth]{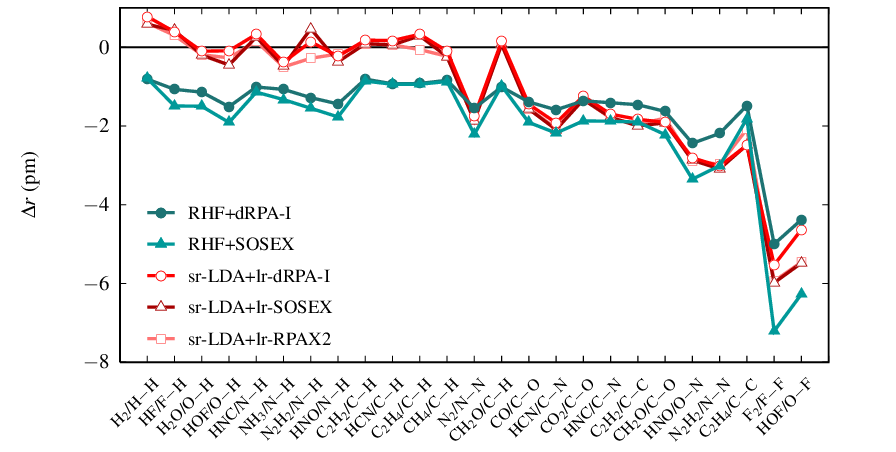}
\captionof{figure}{\label{fig_bond}Deviation of bond distances (pm) of small molecules with respect to the reference~\cite{Pawlowski:02}.}
\end{center}\end{figure}

Table \ref{table_bond} and Figure \ref{fig_bond} show bond lengths and angles obtained from geometry optimizations of a set of small molecules 
(\ce{H2}, HF, \ce{H2O}, HOF, HNC, \ce{NH3}, \ce{N2H2}, HNO, \ce{C2H2}, HCN, \ce{C2H4}, \ce{CH4}, \ce{N2}, \ce{CH2O}, 
CO, {CO2}, \ce{F2})
at the sr-LDA+lr-MP2, sr-LDA+lr-dRPA-I, sr-LDA+lr-SOSEX as well as sr-LDA+lr-RPAX2 levels. 
Optimizations were performed using the program GADGET~\cite{Bucko:05} interfaced with MOLPRO, with the aug-cc-pVQZ basis set, although, as expected in an RSH framework, the convergence of the results is fast with respect to the basis set. 
The calculations were considered to have reached convergence when all gradients components were under 0.0003 a.u.  
Our results are compared to 
the optimized  geometries, recently published by Rekkedal \textit{et al.}~\cite{Rekkedal:13a} at the MP2, dRPA-I and SOSEX levels without range-separation,  
and perfectly reproduced by our present implementation.
The  reference geometries have been taken from the work of Pawlowski \textit{et al.}~\cite{Pawlowski:02} 
and are obtained from experimental rotational constants and vibration-rotation interaction constants computed at the CCSD(T) level with the cc-pVQZ basis. 
The deviations from the reference bond lengths are usually less than 4~pm, except the cases of  \ce{F-F} and \ce{O-F} bond lengths, where the error can be as large as 7~pm. 
In the calculations using RHF orbitals, the mean absolute error (MAE) of the MP2 bond lengths is 0.476 pm while dRPA-I and SOSEX have mean absolute errors of 1.589 pm and 2.077 pm, respectively; the range-hybrid calculations, on the other hand, all yield similar MAE (1.276 pm for sr-LDA+lr-MP2, 1.347 pm for sr-LDA+lr-dRPA-I, 1.459 pm for sr-LDA+lr-SOSEX and 1.388 pm for sr-LDA+lr-RPAX2). 
Among all the methods presented in Table~\ref{table_bond}, it is the full-range MP2 which gives the best results, at least at this relatively high basis set level. A comparison with the MAE values reported by Burow at al.~\cite{Burow:13} indicates that simple PBE and PBE0 functionals have a performance which is not very far from that of full-range MP2. Their dRPA-I calculations using PBE orbitals is slightly better than MP2. It is undeniable that the RSH+RPA version tested here, based on short-range LDA, 
have a worse performance than these above-mentioned methods. 
The range-hybrid sr-LDA+lr-dRPA-I and sr-LDA+lr-SOSEX calculations yield globally better agreement with the reference values then analogous calculations using RHF orbitals. However, it is quite clear that mainly the \ce{X-H} bond lengths are improved in the range-hybrid calculations, while the bond lengths  between first-row atoms remain essentially of similar quality as in RHF+RPA calculations. 
The global agreement is essentially the same between the four range-separated methods (MP2 and dRPA-I, SOSEX, RPAX2) indicating that in these simple cases the bond length corrections, which can be attributed to higher than second order Møller-Plesset effects are negligible. 
This is supported by the fact that different ways of including exchange diagrams in the correlation energy calculations, namely following either SOSEX or RPAX2, does not improve the results significantly. 
The angles are somewhat farther on from the reference in the range-hybrid cases than in the calculations with RHF orbitals, but the differences are only in the order of one degree.

\setlength\tabcolsep{4pt}
\begin{table*}[!htbp]
\footnotesize
\begin{center}
\begin{tabularx}{\linewidth}{lcbbbbbbbb}
\hline\hline
                          &           & \multicolumn{1}{B}{\scriptsize sr-LDA+ lr-MP2} 
                                               & \multicolumn{1}{B}{\scriptsize sr-LDA+ lr-dRPA-I}
                                                        & \multicolumn{1}{B}{\scriptsize sr-LDA+ lr-SOSEX}
                                                                 & \multicolumn{1}{B}{\scriptsize sr-LDA+ lr-RPAX2}
                                                                          & \multicolumn{1}{B}{\scriptsize MP2$^a$}
                                                                                   & \multicolumn{1}{B}{\scriptsize dRPA-I$^a$}
                                                                                            & \multicolumn{1}{B}{\scriptsize SOSEX$^a$}
                                                                                                     & \multicolumn{1}{B}{Ref.~\citep{Pawlowski:02}} \\
\hline\hline                                                                           
H$_2$                     & H$-$H     &  0.572 &  0.783 &  0.581 &  0.616 & -0.543 & -0.814 & -0.794 &  74.149 \\
HF                        & F$-$H     &  0.323 &  0.394 &  0.427 &  0.296 & -0.077 & -1.067 & -1.491 &  91.688 \\
H$_2$O                    & O$-$H     & -0.120 & -0.099 & -0.202 & -0.192 & -0.182 & -1.139 & -1.504 &  95.790 \\
HOF                       & O$-$H     & -0.203 & -0.091 & -0.454 & -0.269 & -0.399 & -1.510 & -1.908 &  96.862 \\
HNC                       & N$-$H     &  0.219 &  0.325 &  0.260 &  0.150 & -0.099 & -1.014 & -1.143 &  99.489 \\
NH$_3$                    & N$-$H     & -0.394 & -0.383 & -0.480 & -0.496 & -0.386 & -1.074 & -1.334 & 101.139 \\
N$_2$H$_2$                & N$-$H     & -0.185 &  0.144 &  0.470 & -0.279 & -0.403 & -1.285 & -1.555 & 102.883 \\
HNO                       & N$-$H     & -0.054 & -0.233 & -0.365 & -0.172 & -0.557 & -1.435 & -1.765 & 105.199 \\
C$_2$H$_2$                & C$-$H     &  0.156 &  0.185 &  0.094 &  0.092 & -0.253 & -0.811 & -0.865 & 106.166 \\
HCN                       & C$-$H     &  0.128 &  0.175 &  0.058 &  0.055 & -0.331 & -0.925 & -0.940 & 106.528 \\
C$_2$H$_4$                & C$-$H     &  0.016 &  0.332 &  0.288 & -0.056 & -0.372 & -0.920 & -0.942 & 108.068 \\
CH$_4$                    & C$-$H     & -0.131 & -0.096 & -0.241 & -0.227 & -0.431 & -0.840 & -0.873 & 108.588 \\
N$_2$                     & N$-$N     & -1.644 & -1.751 & -1.865 & -1.847 &  1.010 & -1.552 & -2.209 & 109.773 \\
CH$_2$O                   & C$-$H     &  0.148 &  0.151 &  0.071 &  0.105 & -0.404 & -1.008 & -0.977 & 110.072 \\
CO                        & C$-$O     & -1.392 & -1.454 & -1.574 & -1.556 &  0.370 & -1.392 & -1.914 & 112.836 \\
HCN                       & C$-$N     & -1.781 & -1.921 & -2.095 & -2.062 &  0.711 & -1.593 & -2.178 & 115.336 \\
CO$_2$                    & C$-$O     & -1.161 & -1.235 & -1.335 & -1.322 &  0.353 & -1.364 & -1.872 & 116.006 \\
HNC                       & C$-$N     & -1.583 & -1.698 & -1.782 & -1.796 &  0.160 & -1.413 & -1.882 & 116.875 \\
C$_2$H$_2$                & C$-$C     & -1.649 & -1.830 & -2.002 & -1.964 &  0.243 & -1.464 & -1.900 & 120.356 \\
CH$_2$O                   & C$-$O     & -1.641 & -1.908 & -1.925 & -1.797 &  0.105 & -1.632 & -2.224 & 120.465 \\
HNO                       & O$-$N     & -2.778 & -2.811 & -2.868 & -2.900 &  0.611 & -2.436 & -3.355 & 120.859 \\
N$_2$H$_2$                & N$-$N     & -2.780 & -3.019 & -3.095 & -2.973 &  0.291 & -2.188 & -3.008 & 124.575 \\
C$_2$H$_4$                & C$-$C     & -1.901 & -2.493 & -2.475 & -2.077 & -0.464 & -1.478 & -1.811 & 133.074 \\
F$_2$                     & F$-$F     & -5.730 & -5.534 & -5.978 & -5.941 & -1.737 & -4.998 & -7.208 & 141.268 \\
HOF                       & O$-$F     & -5.214 & -4.634 & -5.479 & -5.460 & -1.412 & -4.376 & -6.270 & 143.447 \\
HOF                       & H$-$O$-$F &  2.214 & -1.807 &  2.407 &  2.389 &  0.138 &  1.453 &  2.178 &  97.860 \\
H$_2$O                    & H$-$O$-$H &  1.886 &  2.089 &  2.199 &  2.186 & -0.249 &  0.472 &  1.065 & 104.400 \\
N$_2$H$_2$                & H$-$N$-$N &  1.605 &  1.892 &  1.296 &  1.666 & -0.435 &  0.587 &  1.075 & 106.340 \\
NH$_3$                    & H$-$N$-$H &  1.757 &  1.900 &  1.982 &  2.032 & -0.527 & -0.241 &  0.320 & 107.170 \\
HNO                       & H$-$N$-$O &  1.008 &  1.151 &  1.217 &  0.946 & -0.495 &  0.296 &  0.671 & 108.260 \\
C$_2$H$_4$                & C$-$C$-$H &  0.164 & -0.016 &  0.707 &  0.176 & -0.063 &  0.128 &  0.127 & 121.400 \\
CH$_2$O                   & H$-$C$-$O &  0.098 &  0.401 &  0.180 &  0.115 &  0.146 &  0.264 &  0.283 & 121.630 \\
\hline
\multicolumn{2}{l}{$r_e$ MAE}         &  1.276 &  1.347 &  1.459 &  1.388 &  0.476 &  1.589 &  2.077 &         \\
\multicolumn{2}{l}{$r_e$ M\%AE}       &  1.034 &  1.103 &  1.191 &  1.128 &  0.415 &  1.373 &  1.777 &         \\ 
\multicolumn{2}{l}{$\alpha$ MAE}      &  1.247 &  1.322 &  1.427 &  1.359 &  0.293 &  0.492 &  0.817 &         \\
\multicolumn{2}{l}{$\alpha$ M\%AE}    &  1.195 &  1.258 &  1.356 &  1.302 &  0.273 &  0.473 &  0.787 &         \\ 
\hline\hline
\end{tabularx}
\captionof{table}{ \label{table_bond}Deviation with respect to the reference of the bond lengths $r_e$ (pm) and angles $\alpha$ (degrees) of small molecules obtained after geometry optimizations. $^a$from Ref.~\citep{Rekkedal:13a}}
\end{center}
\end{table*}

\begin{figure}[!htbp]
\begin{center}
\includegraphics[width=\linewidth]{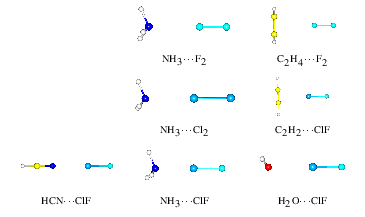}
\captionof{figure}{\label{fig_CT7}Charge transfer complexes of the CT7 set~\cite{Zhao:05a}.}
\end{center}
\end{figure}

\begin{figure}[!htbp]
\begin{center}
\includegraphics[width=\linewidth]{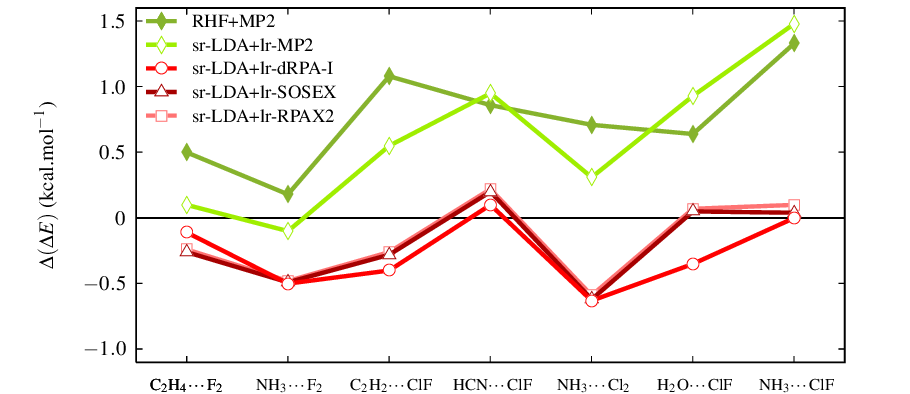}
\captionof{figure}{\label{fig_energy}Deviation of the interaction energies (kcal.mol$^{-1}$) of the CT7 dimers after geometry optimization without counterpoise correction with respect to the reference~\cite{Zhao:05a}.}
\end{center}
\end{figure}

\begin{figure}[!htbp]
\begin{center}
\includegraphics[width=\linewidth]{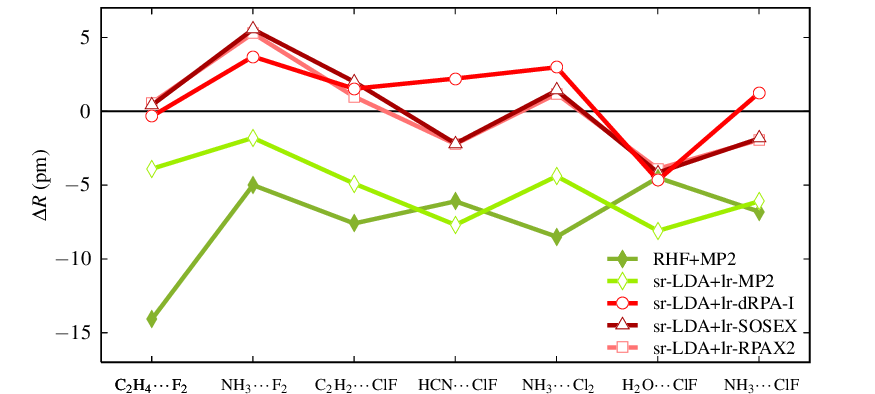}
\captionof{figure}{\label{fig_dist}Deviation of the inter-monomers distances (pm) of CT7 dimers after geometry optimization without counterpoise correction with respect to the reference~\cite{Zhao:05a,tech:Truhlar}.}
\end{center}
\end{figure}

In order to test the gradients for a class of intermolecular interactions, we present interaction energies (Table \ref{table_energy} and Figure \ref{fig_energy}) as well as inter-monomer distances (Table \ref{table_dist} and Figure \ref{fig_dist}) resulting from the geometry optimizations of binary systems in the CT7 (charge transfer) ensemble of intermolecular complexes~\cite{Zhao:05a} at the sr-LDA+lr-dRPA-I, sr-LDA+lr-SOSEX and sr-LDA+lr-RPAX2 levels.
For an illustration of the relative orientation of the monomers, see Figure \ref{fig_CT7}. 
Our results are compared to those from Chabbal \textit{et al.}~\cite{Chabbal:10} obtained by geometry optimizations at the MP2 and sr-LDA+lr-MP2 levels and to reference values given by the group of Truhlar~\cite{Zhao:05a,tech:Truhlar}. 
The optimizations were conducted with GADGET program using the aug-cc-pVTZ basis set, without counterpoise correction. It is believed that for range-hybrid calculations the effect of the basis set superposition error is small, even at this relatively modest basis set level.

These charge transfer complexes are usually problematic for plain DFT methods due to the sizable delocalization error of the common functionals, as it has been well-known for a long time~\cite{Ruiz:96}. Therefore it is expected that the RSH determinant is going to be a reasonable reference state for the long-range RPA correlation calculations to take into account the London dispersion interactions stabilizing these complexes.
The study from Chabbal \textit{et al.} has already shown that the range-separated MP2 approach improves the results with respect to full-range MP2 for charge transfer complexes.
Our results demonstrate that the three lr-RPA variants tested yield a general improvement over RSH+MP2 calculations, with mean absolute errors around 0.30 kcal.mol$^{-1}$ for the interaction energies and around 2.5 pm for the inter-monomer distances. 

The percentage errors of all methods for both the interaction energy and the inter-monomer distance  is rather high in the case of the \ce{NH3...F2} dimer (c.f.\ for example the visual abstract which shows the percentage deviation of the inter-monomer distances). This observation can be attributed to the relatively small magnitude of the reference values, especially in the case of inter-monomer distances. 
While the \ce{H2O...ClF} interaction energy is underestimated by sr-LDA+lr-dRPA-I and well recovered by sr-LDA+lr-SOSEX and sr-LDA+lr-RPAX2, the inter-monomer distance after geometry optimization is consistently less good using any of the range-hybrid RPA method, with errors around 4.5~pm. 
On the contrary, while the inter-monomer distance of the \ce{NH3...Cl2} dimer is close to the reference for all the range-hybrid RPA methods, the interaction energies show the largest deviations (around 0.6 kcal.mol$^{-1}$) in the whole set of systems. 
We are going to attempt a rationalization of these observations in the next paragraph.

\begin{table*}[!htbp]
\footnotesize
\begin{center}
\begin{tabularx}{\linewidth}{Bsbbbbs}
\hline\hline
                         & \multicolumn{1}{S}{MP2$^a$}
                                   & \multicolumn{1}{B}{sr-LDA+ lr-MP2$^a$}
                                                 & \multicolumn{1}{B}{sr-LDA+ lr-dRPA-I}
                                                              & \multicolumn{1}{B}{sr-LDA+ lr-SOSEX}
                                                                           & \multicolumn{1}{B}{sr-LDA+ lr-RPAX2}
                                                                                     & \multicolumn{1}{S}{Ref.~\citep{Zhao:05a}}   \\ 
\hline\hline
C$_2$H$_4$$\cdots$F$_2$  &  1.56   &  1.16       &   0.95     &  0.80      &  0.82   &  1.06                  \\
    NH$_3$$\cdots$F$_2$  &  1.99   &  1.71       &   1.31     &  1.32      &  1.33   &  1.81                  \\
C$_2$H$_2$$\cdots$ClF    &  4.89   &  4.36       &   3.41     &  3.53      &  3.55   &  3.81                  \\
       HCN$\cdots$ClF    &  5.72   &  5.81       &   4.96     &  5.06      &  5.08   &  4.86                  \\
    NH$_3$$\cdots$Cl$_2$ &  5.59   &  5.19       &   4.25     &  4.26      &  4.29   &  4.88                  \\
    H$_2$O$\cdots$ClF    &  6.00   &  6.29       &   5.01     &  5.41      &  5.43   &  5.36                  \\
    NH$_3$$\cdots$ClF    & 11.95   & 12.10       &  10.62     & 10.66      & 10.72   & 10.62                  \\
\hline
\raggedright MAE         &  0.76   &  0.63       &   0.30     &  0.28      &  0.28   &                        \\
\raggedright M\%AE       & 20.31   & 12.37       &  10.05     & 10.98      & 10.70   &                        \\
\hline\hline
\end{tabularx}
\caption{\label{table_energy}
Interaction energies (kcal.mol$^{-1}$) of the CT7 dimers after geometry optimization without counterpoise correction. 
                             $^a$results from~\cite{Chabbal:10}.}
\end{center}
\end{table*}

\begin{figure}[!htbp]
\begin{center}
\includegraphics[width=\linewidth]{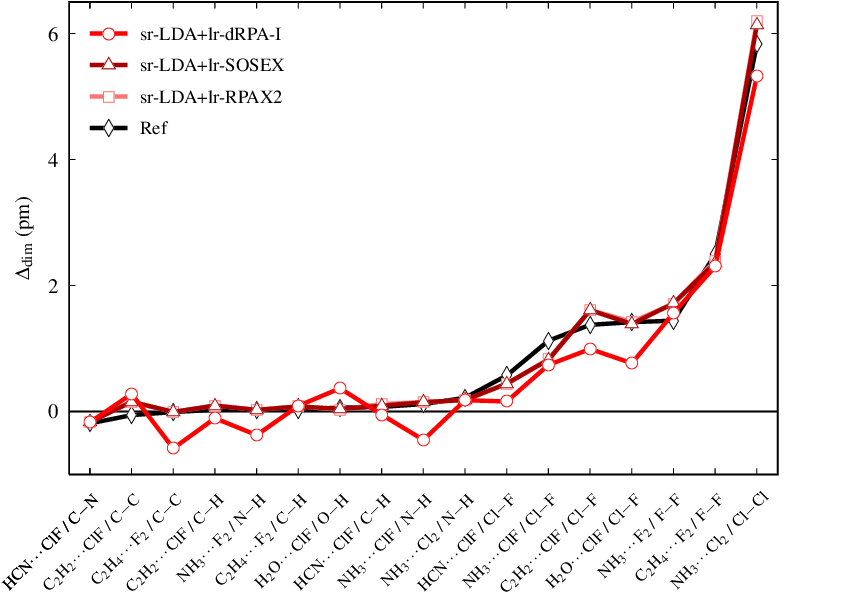}
\captionof{figure}{\label{fig_dim}Effect of the dimerization on the bond lengths (pm) of the monomers of the CT7 dimers. The reference is from~\cite{Zhao:05a,tech:Truhlar}.}
\end{center}
\end{figure}

\begin{table*}[!htbp]
\footnotesize
\begin{center}
\begin{tabularx}{\linewidth}{Bsbbbbs}
\hline\hline
                         & \multicolumn{1}{S}{MP2$^a$}
                                   & \multicolumn{1}{B}{sr-LDA+ lr-MP2$^a$}
                                                 & \multicolumn{1}{B}{sr-LDA+ lr-dRPA-I}
                                                              & \multicolumn{1}{B}{sr-LDA+ lr-SOSEX}
                                                                           & \multicolumn{1}{B}{sr-LDA+ lr-RPAX2}
                                                                                     & \multicolumn{1}{S}{Ref.~\citep{Zhao:05a,tech:Truhlar}}   \\
\hline\hline
C$_2$H$_4$$\cdots$F$_2$  & 291.2   & 301.4       & 305.0      & 305.7      &  305.9  & 305.3                     \\
    NH$_3$$\cdots$F$_2$  & 264.6   & 267.8       & 273.3      & 275.2      &  274.9  & 269.6                     \\
C$_2$H$_2$$\cdots$ClF    & 280.0   & 282.7       & 289.1      & 289.6      &  288.6  & 287.6                     \\
       HCN$\cdots$ClF    & 254.8   & 253.2       & 263.1      & 258.7      &  258.7  & 260.9                     \\
    NH$_3$$\cdots$Cl$_2$ & 260.3   & 264.4       & 271.8      & 270.2      &  270.0  & 268.8                     \\
    H$_2$O$\cdots$ClF    & 251.2   & 247.6       & 251.0      & 251.5      &  251.8  & 255.7                     \\
    NH$_3$$\cdots$ClF    & 223.4   & 224.1       & 231.5      & 228.4      &  228.3  & 230.2                     \\
\hline                                                                              
\raggedright MAE         &   7.5   &   5.3       &   2.4      &   2.5      &    2.3  &                           \\
\raggedright M\%AE       &   2.8   &   2.0       &   0.9      &   1.0      &    0.9  &                           \\
\hline\hline
\end{tabularx}
\caption{\label{table_dist}
Inter-monomers distances (pm) of CT7 dimers after geometry optimization without counterpoise correction. 
                           $^a$results from~\cite{Chabbal:10}.}
\end{center}
\end{table*}

We propose on Table \ref{table_dim} and Figure \ref{fig_dim} an analysis of the bond lengths obtained after geometry optimizations \textit{via} the difference $\Delta_\T{dim}=r_\T{dimer}\!-\!r_\T{mono}$ between the bond lengths optimized in the dimer ($r_\T{dimer}$) and the  independently optimized bond lengths in the monomers ($r_\T{mono}$). This quantity measures the effect of the dimerization on the geometry of the monomers. 
The $r_\T{dimer}$ and $\Delta_\T{dim}$ values resulting from sr-LDA+lr-dRPA-I, sr-LDA+lr-SOSEX and sr-LDA+lr-RPAX2 calculations and for the reference geometries from Zhao and Truhlar~\cite{Zhao:05a,tech:Truhlar} obtained at the MC-QCISD/3 level, are shown in Table~\ref{table_dim}. 
The bonds are collected in different groups: \ce{C-X}, \ce{X-H}, \ce{X-X} (heteroatomic and homoatomic). 
We observe that the bonds are generally slightly less deformed by the intermolecular interactions in the case of the sr-LDA+lr-dRPA-I calculations than in reference geometries, and are better recovered at the sr-LDA+lr-SOSEX and sr-LDA+lr-RPAX2 levels (see Figure \ref{fig_dim}). 
Both bonds involved in the dimer \ce{NH3...F2}  reflect the standard behavior of their groups.  This confirms that this complex is not an inherently problematic case in range-hybrid RPA. 
The \ce{NH3...Cl2} complex shows internal bond lengths far off as compared to the standard behavior. The bonds are more deformed in the RSH-RPA calculations in comparison to the reference geometry, which explains a large underestimation of the interaction energy in spite of the good inter-monomer distances.
We see that the bonds involved in the \ce{H2O...ClF} dimer, which showed bad inter-monomer distances for all RSH-RPAs, are described much better by the sr-LDA+lr-SOSEX and sr-LDA+lr-RPAX2 methods than by the sr-LDA+lr-dRPA-I: this could explain the improvement previously mentioned for the interaction energy of this dimer.

\begin{landscape}
\begin{table}
\footnotesize
\begin{center}
\begin{tabularx}{1.3\textwidth}{cc|rrr|rrr|rrr|rr}
\hline\hline
                         &         & \multicolumn{3}{c|}{sr-LDA+lr-dRPA-I}                           
                                                                & \multicolumn{3}{c|}{sr-LDA+lr-SOSEX}                           
                                                                                            & \multicolumn{3}{c|}{sr-LDA+lr-RPAX2}                           
                                                                                                                        & \multicolumn{2}{c}{Ref.~\cite{Zhao:05a,tech:Truhlar}}   \\
                         &         & \multicolumn{1}{B}{$r_\T{dimer}$}
                                              & \multicolumn{1}{B}{$\Delta_\T{dim}$}
                                                       & \multicolumn{1}{B|}{$\Delta \Delta$}
                                                                & \multicolumn{1}{B}{$r_\T{dimer}$}
                                                                          & \multicolumn{1}{B}{$\Delta_\T{dim}$}
                                                                                   & \multicolumn{1}{B|}{$\Delta \Delta$}
                                                                                             & \multicolumn{1}{B}{$r_\T{dimer}$}
                                                                                                       & \multicolumn{1}{B}{$\Delta_\T{dim}$}
                                                                                                                & \multicolumn{1}{B|}{$\Delta \Delta$}
                                                                                                                         & \multicolumn{1}{B}{$r_\T{dimer}$}
                                                                                                                                   & \multicolumn{1}{B}{$\Delta_\T{dim}$ }\\
\hline\hline
       HCN$\cdots$ClF    &  C$-$N  &  113.363 & -0.173 &  0.016 &  113.171 & -0.166 &  0.023 & 113.206 & -0.166 &  0.023 & 115.621 & -0.189 \\
C$_2$H$_2$$\cdots$ClF    &  C$-$C  &  118.788 &  0.179 & -0.039 &  118.613 &  0.182 & -0.036 & 118.656 &  0.181 & -0.037 & 121.058 &  0.218 \\
C$_2$H$_4$$\cdots$F$_2$  &  C$-$C  &  131.237 &  0.368 &  0.306 &  131.025 &  0.042 & -0.020 & 131.050 &  0.029 & -0.033 & 133.678 &  0.062 \\[0.5em]
C$_2$H$_2$$\cdots$ClF    &  C$-$H  &  106.321 & -0.051 & -0.123 &  106.362 &  0.090 &  0.017 & 106.400 &  0.120 &  0.047 & 106.668 &  0.073 \\
    NH$_3$$\cdots$F$_2$  &  N$-$H  &  100.465 & -0.375 & -0.405 &  100.741 &  0.025 & -0.005 & 100.753 &  0.025 & -0.005 & 101.524 &  0.030 \\
C$_2$H$_4$$\cdots$F$_2$  &  C$-$H  &  108.111 & -0.585 & -0.576 &  108.045 & -0.006 &  0.003 & 108.054 & -0.008 &  0.000 & 108.437 & -0.009 \\
    H$_2$O$\cdots$ClF    &  O$-$H  &   95.356 & -0.452 & -0.570 &   95.853 &  0.149 &  0.031 &  95.861 &  0.147 &  0.028 &  96.301 &  0.118 \\
       HCN$\cdots$ClF    &  C$-$H  &  106.658 & -0.107 & -0.135 &  106.692 &  0.094 &  0.066 & 106.699 &  0.094 &  0.066 & 106.930 &  0.028 \\
    NH$_3$$\cdots$ClF    &  N$-$H  &  101.121 &  0.280 &  0.339 &  100.870 &  0.154 &  0.212 & 100.882 &  0.153 &  0.212 & 101.435 & -0.058 \\
    NH$_3$$\cdots$Cl$_2$ &  N$-$H  &  100.931 &  0.090 &  0.058 &  100.797 &  0.081 &  0.050 & 100.810 &  0.081 &  0.050 & 101.525 &  0.031 \\[0.5em]
       HCN$\cdots$ClF    & Cl$-$F  &  162.386 &  0.992 & -0.384 &  162.263 &  1.608 &  0.232 & 162.320 &  1.612 &  0.236 & 165.635 &  1.376 \\
    NH$_3$$\cdots$ClF    & Cl$-$F  &  166.718 &  5.325 & -0.511 &  166.796 &  6.140 &  0.304 & 166.901 &  6.193 &  0.357 & 170.095 &  5.836 \\
C$_2$H$_2$$\cdots$ClF    & Cl$-$F  &  162.159 &  0.766 & -0.652 &  162.044 &  1.389 & -0.029 & 162.131 &  1.422 &  0.004 & 165.677 &  1.418 \\
    H$_2$O$\cdots$ClF    & Cl$-$F  &  162.956 &  1.563 &  0.121 &  162.372 &  1.717 &  0.275 & 162.419 &  1.711 &  0.269 & 165.701 &  1.442 \\[0.5em]
    NH$_3$$\cdots$F$_2$  &  F$-$F  &  136.602 &  0.740 & -0.384 &  136.226 &  0.821 & -0.303 & 136.295 &  0.833 & -0.291 & 142.547 &  1.124 \\
C$_2$H$_4$$\cdots$F$_2$  &  F$-$F  &  136.020 &  0.159 & -0.414 &  135.849 &  0.443 & -0.130 & 135.900 &  0.438 & -0.136 & 141.996 &  0.573 \\
    NH$_3$$\cdots$Cl$_2$ & Cl$-$Cl &  199.898 &  2.303 & -0.193 &  199.388 &  2.363 & -0.133 & 199.526 &  2.395 & -0.101 & 203.628 &  2.496 \\
\hline\hline
\end{tabularx}
\caption{\label{table_dim} Data concerning the bond lengths (pm) of the monomers of the CT7 dimers. 
$r_\T{dimer}$ are the bond lengths optimized in the dimer. 
$\Delta_\T{dim}$ is the effect of the dimerization on the bond lengths 
  ($\Delta_\T{dim}>0$ shows a bond that is longer in the dimer), 
$\Delta \Delta=\Delta_\T{dim}^\T{method} - \Delta_\T{dim}^\T{ref}$ is the compared effect of the dimerization between a given method's level and the MC-QCISD/3 level.
}
\end{center}
\end{table}
\end{landscape}

\section{Conclusions and outlook}

The RSH+RPA analytical energy gradients have been derived using the Lagrangian formulation and implemented in the development version of the MOLPRO quantum chemical program package. 
Although the working expressions have been obtained for all of the main categories of the RPA correlation energy, the present work reports the numerical implementation only for the direct RPA, the SOSEX and the RPAX2 variants. 
These results show a significant improvement with respect to range-separated MP2 calculations in the description of both the energetics and the structure of charge transfer complexes, where intermolecular interactions play an important role even in the geometry of the constituents of the complexes.

The present numerical implementation provides mainly reference data for relatively small systems. Admittedly, the computational efficiency of an orbital-based algorithm, used here, is quite limited. However, generalizations for density-fitting, resolution-of-identity and even Cholesky decomposition algorithms (outlined e.g.in Ref.~\cite{Scuseria:08}) seem to be rather straightforward and will be the subject of future work. 
Another extension of the present work consists in the computational realization of the exchange-including rCCD/RPAx correlation energy expressions, namely the SO1 and SO2 variants~\cite{Toulouse:11}, which have shown the best qualitative performance in range-hybrid calculations of intermolecular interaction energies. 

It is quite clear from our past experience, that for simple intermolecular interact action energies the short-range functional has a relatively minor influence on the quality of the results. The situation seems to be different as far as we would like to reproduce bond lengths and angles, and the use of  short-range GGA (e.g.\sr-PBE) functionals is mandatory  to improve these results. Note that in geometry optimizations, in addition to the sr exchange-correlation functionals and potentials one needs sr-PBE kernels (second functional derivatives of the sr-PBE functionals). Work in this direction is in progress and we hope to test this hypothesis in the near future. 

As a by-product of the analytical force implementation, the non-relaxed and relaxed density matrices at the RSH-RPA levels are available. They will be exploited for the analysis of the correlation effects on one-electron properties, like charge densities and their multipole moments. A study in this direction is in preparation.

\begin{acknowledgement}
B.M. is grateful for the hospitality of the Chemistry Department of the 
E\"otv\"os Lor\'and University Budapest during his stay and for the financial support provided by the
T\'AMOP, supported by the European Union and co-financed by the European Social Fund (Grant Agreement No. T\'AMOP 4.2.1/B-09/1/KMR- 2010-0003). 
J.G.A. is grateful to the CEU-IAS (Budapest) for the senior fellowship during the academic year 2011/2012. P.G.S. acknowledges financial support by Országos Tudományos Kutatási Alap (OTKA); contract grant number: F104672. 
\end{acknowledgement}

\appendix    

\section{Stationary conditions with respect to orbital coefficients}
\label{appsec:statioC}

Let us parameterize the variation of orbital coefficients at first order by a unitary rotation matrix $\b{V}$ as $\b{C}\leftarrow\b{C}+\b{C}\b{V}$. The stationary conditions for the Lagrangian can be written as

\begin{align}
\half\left.\derivR{\mathcal{L}}{\b{V}}\right|_{\b{V}=\b{0}} = 0
\end{align}

The factor of $\Shalf$ is inserted to compensate a factor of 2 appearing in the upcoming derivations for reasons of symmetry. 

It will prove to be convenient to rewrite the Lagrangian of Eq.~(\ref{eq:lag}) by factorizing the terms which depend on the orbital coefficients. This can be achieved by separating in $E\RPA\lr(\b{C},\b{T}) + \trace{\B{\ll}\b{R}(\b{C},\b{T})}$ the terms depending on the fockian from those depending on the two-electron integrals, leading to:

\begin{align}
\mathcal{L}(\b{C},\b{T},\b{z},\b{x},\B{\ll})&=
  \trace{\b{d}\b{f}} + \DC
      \nonumber\\&\quad         
+ \trace{\b{M}\b{K}} + \trace{\b{N}\b{K}^\prime}
      \nonumber\\&\quad         
+ \trace{\b{x}\left(\b{C}^\trnp\b{S}\b{C}-\b{1}\right)},
\label{appeq:lagFac}
\end{align}
where the super-matrices $\b{M}$ and $\b{N}$ gather all the elements that are multiplied by the integrals $\b{K}$ and $\b{K}^\prime$, respectively. The particular forms of these super-matrices thus depend on the Riccati equation and on the energy expression corresponding to the given RPA variant chosen for the long-range correlation energy. 
We defined the relaxed density $\b{d}=\b{d}\zero+\b{d}\deux+\b{z}$ with the matrix $\b{d}\deux$, whose blocks are:

\begin{align}
\left(\b{d}\deux\right)_{ij} &
                               =-\FdTr{\b{T}}{\B{\lambda}}_{ij} - \FdTr{\B{\lambda}}{\b{ T}}_{ij} \nonumber\\
\left(\b{d}\deux\right)_{ab} &
                               = \FdTr{\b{T}}{\B{\lambda}}_{ab} + \FdTr{\B{\lambda}}{\b{ T}}_{ab}          \\
\left(\b{d}\deux\right)_{ai} &= 0                                                                 \nonumber
\end{align}

In the above equations we use a specific notation for the "failed traces", that is to say for the partial summations leading to a result which still depends on two of the four indexes that compose the super-indexes of the super-matrices:
 
\begin{align}
\FdTr{\b{X}}{\b{Y}}_{ij}&=\sum_{kc,a} X_{\b{i}a,kc}Y_{kc,\b{j}a} \\
\FdTr{\b{X}}{\b{Y}}_{ab}&=\sum_{kc,i} X_{i\b{a},kc}Y_{kc,i\b{b}}
\end{align}

The derivatives of all the terms in Eq.~(\ref{appeq:lagFac}) with respect to a change in the orbital coefficient are fairly lengthy, therefore only some of the elements are given here. From the derivation of the trace of the two-electron integrals with the super-matrix $\b{M}$ will emerge contractions of the form 
    $\FdTr{\b{K}}           {\b{M}}_{ij}$ 
and $\FdTr{\b{K}}           {\b{M}}_{ab}$ 
as well as generalizations of the form 
    $\FdTr{\overline{\b{K}}}{\b{M}}_{ia}$ 
and $\FdTr{\overline{\b{K}}}{\b{M}}_{ai}$ where the super-matrix $\overline{\b{K}}$ is constructed from $\b{K}$ but does not respect it's $ia,jb$ structure (the same quantities are derived for the trace with $\b{N}$). All those terms are grouped in the matrix $\B{\Theta}$.

\section{Two-electron fockian and double-count derivatives}

The derivatives of the long- and short-range parts of the "fockian plus double-count" terms show some interesting analogies. The derivative of the long-range two-electron contribution $\trace{\b{d}\b{g}\lr\left[\b{d}\zero\right]}$ yields a $\b{g}\lr\left[\b{d}\zero\right]\b{d}$, and, by an interchange property of the indexes involved in the summations, $\b{g}\lr\left[\b{d}\right]\b{d}\zero$. The derivation of the double-count correction will cancel out the $\b{g}\lr\left[\b{d}\zero\right]\b{d}\zero$ in the "interchanged" term, so that we finally obtain:

\begin{align}
&\half\left.\derivR{}{\b{V}}\left(\trace{\b{d}\b{g}\lr}+\DC\lr\right)\right|_{\b{V}=\b{0}}
\nonumber\\&\qquad\qquad
=\b{g}\lr\b{d}+\b{g}\lr\left[\b{d}\deux+\b{z}\right]\b{d}\zero
\label{appeq:LRstatio}
\end{align}

In a similar, but less obvious manner, the derivative of the short-range contribution $\trace{\b{d}\b{g}\sr}$ gives $\b{g}\sr\b{d}$ and, by a comparable phenomenon, a new object that we call $\b{w}\sr\left[\b{d}\right]\b{d}\zero$. Using the relationship:

\begin{align}
\half\left.\derivR{E\Hxc\sr[n_0]}{\b{V}}\right|_{\b{V}=\b{0}}
=\b{g}\sr\b{d}\zero,
\label{appeq:relation}
\end{align}
(see Appendix \ref{appsec:derivSR}) we see that the double-count term behaves as follow:

\begin{align}
&\half\left.\derivR{}{\b{V}}\right|_{\b{V}=\b{0}}\left(E\Hxc\sr[n_0]-\b{d}\zero\b{g}\sr\right)
\nonumber\\&\qquad\qquad
=\b{g}\sr\b{d}\zero
-\b{g}\sr\b{d}\zero-\b{w}\sr\left[\b{d}\zero\right]\b{d}\zero,
\end{align}
and cancels out the $\b{w}\sr\left[\b{d}\zero\right]\b{d}\zero$ from the "interchanged" term, much like previously the long-range double-count term cancelled out $\b{g}\lr\left[\b{d}\zero\right]\b{d}\zero$ from the long-range "interchanged" term. This leads to the result:

\begin{align}
&\half\left.\derivR{}{\b{V}}\left(\trace{\b{d}\b{g}\sr}+\DC\sr\right)\right|_{\b{V}=\b{0}}
\nonumber\\&\qquad\qquad
=\b{g}\sr\b{d}+\b{w}\sr\left[\b{d}\deux+\b{z}\right]\b{d}\zero
\label{appeq:SRstatio}
\end{align}

The relationships (\ref{appeq:LRstatio}) and (\ref{appeq:SRstatio}) are similar to each other and their terms enter in the definition of $\B{\Theta}$ and $\tilde{\B{\Theta}}(\b{z})$.

\section{Elaboration on the sr-DFT terms}
\label{appsec:derivSR}

In order to derive some of the quantities needed for the short-range treatment of the gradient, we write the Hartree-exchange-correlation functional as (cf. Eq.~(\ref{eq:Hxc})):

\begin{equation}
E\Hxc\sr[n_\Phi]=\int dr\; F\bigl(\B{\xi}(r)\bigr),
\end{equation}
where $\B{\xi}$ is an array of quantities that enter in the definition of the functional, \textit{i.e.} $\B{\xi}=\{\xi_A\}=\{n_\alpha, \nabla n_\alpha, \nabla n_\beta, \dots \}$.  
For all such objects, that do not include explicitly virtual orbitals, we can write in a most general way:

\begin{align}
\xi_A=\prod_n \fct{n}{\xi_A}{n_0} = \prod_n \sum_{pq} \fct{n}{\xi_A}{d\zero_{pq}\phi_p^\dagger\phi_q},
\end{align}
where $\mathcal{X}_n^{\xi_A}$ are functions that are different for every quantity $\xi_A$. 

Initially, with these notations at hand, we are going to show that the derivative of the functional with respect to the orbital rotation parameters is related to the two-electron part of the short-range fockian, \textit{i.e.}~we are to going to prove Eq.~(\ref{appeq:relation}). The derivative of the functional with respect to the orbital rotation parameters reads:

\begin{align}
\half\left.\derivR{E\Hxc\sr}{V_{rs}}\right|_{V_{rs}=0}=
\half \sum_A \int dr\; \derivR{F}{\xi_A}\left.\derivR{\xi_A}{V_{rs}}\right|_{V_{rs}=0}
\label{appeq:wrtV0}
\end{align}
and the two-electron part of the short-range fockian is:

\begin{align}
g\sr_{rs}=
\sum_A \int dr\; \derivR{F}{\xi_A}\derivR{\xi_A}{d\zero_{rs}}
\label{appeq:wrtC0}
\end{align}

The derivative of a quantity $\xi_A$ with respect to a change of the orbital coefficients, appearing in Eq.~(\ref{appeq:wrtV0}), is 

\begin{align}
\left.\derivR{\xi_A}{V_{rs}}\right|_{V_{rs}=0}
&= \sum_i \left\{\sum_{pq} \fct{i}{\xi_A}{d\zero_{pq}\left(\delta_{ps}\phi_r^\dagger \phi_q+\delta_{qs}\phi_p^\dagger \phi_r\right)}\right\}\prod_{n\neq i}\fct{n}{\xi_A}{n_0}
\nonumber\\
&=2\sum_p\left(\sum_i \fct{i}{\xi_A}{\phi_r^\dagger \phi_p}\prod_{n\neq i}\fct{n}{\xi_A}{n_0}\right)d\zero_{ps},
\label{appeq:wrtV}
\end{align}
while the derivative with respect to $\b{d}\zero$, which occur in Eq.~(\ref{appeq:wrtC0}) is 

\begin{align}
\derivR{\xi_A}{d\zero_{rs}}=
 \sum_i \fct{i}{\xi_A}{\phi_r^\dagger \phi_s}           \prod_{n\neq i}\fct{n}{\xi_A}{n_0}. 
\end{align}
The above results allow us to write:

\begin{align}
\half\left.\derivR{E\Hxc\sr}{V_{rs}}\right|_{V_{rs}=0}
&=\sum_A \int dr\; \derivR{F}{\xi_A}\,\times
\nonumber\\&\qquad
\sum_p
\left(\sum_i \fct{i}{\xi_A}{\phi_r^\dagger \phi_p}\prod_{n\neq i}\fct{n}{\xi_A}{n_0}\right)d\zero_{ps}
\nonumber\\
&=\sum_A \int dr\; \derivR{F}{\xi_A}\sum_p\derivR{\xi_A}{d\zero_{rp}}d\zero_{ps}
\nonumber\\
&=\left(\b{g}\sr\b{d}\zero\right)_{rs}
\end{align}

Subsequently, we will show that the derivative of the trace of the short-range fockian, found in Eq.~(\ref{appeq:SRstatio}), is indeed the sum of terms $\b{g}\sr\b{d}$ and $\b{w}\sr\left[\b{d}\right]\b{d}\zero$. 
The derivative reads:

\begin{multline}
 \sum_{AB} \half\int dr\; \derivR{^2 F}{\xi_B\xi_A}\left.\derivR{\xi_B}{V_{rs}}\right|_{V_{rs}=0}\sum_{pq}\derivR{\xi_A}{d\zero_{pq}}d_{pq}
\\
+\sum_A    \half\int dr\; \derivR{   F}{     \xi_A}\left.\derivR{     }{V_{rs}}\left(\sum_{pq}\derivR{\xi_A}{d\zero_{pq}}d_{pq}\right)\right|_{V_{rs}=0}
\label{appeq:derivDF}
\end{multline}

Inspection of the element ${\textstyle \sum_{pq} \derivR{\xi_A}{d\zero_{pq}}d_{pq}}$ occurring in both terms of Eq.~(\ref{appeq:derivDF}) reveals that it is a sum of objects $\xi_A$ where \textit{one} occurence of $\b{d}\zero$ has been eliminated by the derivation and replaced by the density matrix $\b{d}$. We call this element $\xi_A^{\b{d}}$, it's expression is:

\begin{align}
\xi_A^{\b{d}}=\sum_{pq}\derivR{\xi_A}{d\zero_{pq}}d_{pq}=\sum_i \fct{i}{\xi_A}{n_{\b{d}}}                   \prod_{n\neq i}\fct{n}{\xi_A}{n_0},
\end{align}

The first term of Eq.~(\ref{appeq:derivDF}) involves only elements of the form "$\sqcup_{rp}d\zero_{ps}$" (see Eq.~(\ref{appeq:wrtV})) that enter in the composition of $w\sr\left[\b{d}\right]_{rp}d\zero_{ps}$. 
The derivation of the second term in Eq.~(\ref{appeq:derivDF}) is more involved. 
Two types of terms will arise from the derivative with respect to $\b{V}$ of the 
element $\xi_A^{\b{d}}$
, as follows: 

\begin{align}
         \left.\derivR{\xi_A^{\b{d}}}{V_{rs}}                                                                           \right|_{V_{rs}=0}
&=       \left.\derivR{           }{V_{rs}}\left(\sum_i \fct{i}{\xi_A}{n_{\b{d}}}\prod_{n\neq i}\fct{n}{\xi_A}{n_0}\right)\right|_{V_{rs}=0}
\nonumber\\
&=      \sum_i                                \left.\derivR{\fct{i}{\xi_A}{n_{\b{d}}}}{V_{rs}}\right|_{V_{rs}=0}\prod_{          n\neq i          }\fct{n}{\xi_A}{n_0}
\nonumber\\
&\quad +\sum_i \fct{i}{\xi_A}{n_{\b{d}}} \sum_j \left.\derivR{\fct{j}{\xi_A}{n_{0    }}}{V_{rs}}\right|_{V_{rs}=0}\prod_{\substack{n\neq i\\n\neq j}}\fct{n}{\xi_A}{n_0}
\end{align}

The first type of terms involves the derivative with respect to a rotation of the orbital coefficients of a $\mathcal{X}^{\xi_A}$ that has been "contaminated" by $\b{d}$: they raise elements of the kind "$\sqcup_{rp}d_{ps}$" that contribute to $\b{g}\sr\b{d}$; the second type of terms involves the derivative with respect to $\b{V}$ of an "original" $\mathcal{X}^{\xi_A}$ and will see the emergence of elements "$\sqcup_{rp}d\zero_{ps}$" that compose $\b{w}\sr\left[\b{d}\right]\b{d}\zero$. 

After all the derivations described above have been carried out, the full definition of $\b{w}\sr\left[\b{d}\right]$ is:

\begin{multline}
w\sr[\b{d}]_{pq}
=\sum_A \int dr\; \sum_{ij}\Big( \sum_B\derivR{^2 F}{\xi_A\xi_B}\xi_B^\b{d}\fct{j}{\xi_A}{n_0}
\\
                                 +\derivR{F   }{\xi_A     }           \fct{j}{\xi_A}{n_\b{d}}
                           \Big) \fct{i}{\xi_A}{\phi^\dagger_p\phi_q}\prod_{\substack{n\neq i\\n\neq j}}\fct{n}{\xi_A}{n_0}
\end{multline}

With the notations introduced here, the derivative of the short-range fockian term with respect to the atomic coordinates that appear in Eq.~(\ref{eq:grad}) is derived as:

\begin{align}
E_\text{DFT}\srx&=E_{Hxc}\srx+\trace{\left(\b{d}\deux+\b{z}\right) \b{g}\srx}
\nonumber\\
        &=\sum_A\left\{ \int                 F(\xi_A(r)) + \derivR{F}{\xi_A}\sum_{pq}\derivR{\xi_A(r)}{d\zero_{pq}}\left(\b{d}\deux+\b{z}\right)_{qp}\right\}\x
\end{align}

With a quadrature on a grid of the real space of points $\{\lambda\}$ of weights $\omega_\lambda$, this leads to Eq.~\ref{eq:gradsrdft} of the main text.

%
%


\newpage

\bibliography{MusSzaAng-2014}

\end{document}